\documentstyle[aps,prb,epsf,multicol]{revtex}
\begin{document}

\title{Charge--order transition in the extended Hubbard model
       on a two--leg ladder}
\author{Matthias Vojta$^{(a)}$, R.~E. Hetzel$^{(b)}$, and R.~M. Noack$^{(c)}$}
\address{
  (a) Department of Physics, Yale University,
      New Haven, CT 06520-8120, USA \\
  (b) Institut f\"{u}r Theoretische Physik,
      Technische Universit\"{a}t Dresden, D-01062 Dresden, Germany \\
  (c) Institut de Physique Th\'eorique, Universit\'e de Fribourg,
      CH-1700 Fribourg, Switzerland
}
\date{\today}
\maketitle

\begin{abstract}
We investigate the charge-order transition at zero temperature
in a two-leg Hubbard ladder with additional nearest-neighbor
Coulomb repulsion $V$ using the Density Matrix Renormalization Group
technique.
We consider electron densities between quarter and half filling.
For quarter filling and $U=8 t$, we find evidence
for a
continuous phase transition between a homogeneous state
at small $V$ and a broken-symmetry state with ``checkerboard''
[wavevector ${\bf Q}=(\pi,\pi)$] charge order at large $V$.
This transition to a checkerboard charge-ordered state remains present at
all larger fillings, but becomes discontinuous at sufficiently large
filling.
We discuss the influence of $U/t$ on the transition
and estimate the position of the tricritical points.

\end{abstract}


\widetext
\begin{multicols}{2}
\narrowtext



The competition between kinetic energy and Coulomb repulsion in
electronic systems can lead to a variety of interesting phenomena,
one of them being charge ordering.
A periodic charge order, i.e., a modulation of the charge density,
can be described as charge density wave.
One possible mechanism for charge ordering is the crystallization of
electrons due to their long-range Coulomb repulsion as proposed by
Wigner. \cite{Wigner}
The Wigner lattice of electrons forms {\it without} the underlying lattice
structure
at low densities when the system is dominated by the effect of the
Coulomb repulsion.
Charge ordering may also occur at higher electron densities
if the kinetic energy is reduced due to small hybridization of orbitals,
or due to the interaction with lattice or spin degrees of freedom.
Charge-ordered states have been observed in, for example, rare
earth manganites, which have attracted attention recently due to their
``colossal'' magnetoresistance.
Several of these compounds (e.g. La$_{1-x}$Ca$_x$MnO$_3$ for
$x\geq 0.5$) show a charge-ordered ground state for a
certain range of doping.\cite{Manganites}
Another material showing charge order is NaV$_2$O$_5$.
It undergoes a phase transition at $T_c=34$ K that is characterized by
the opening of a spin gap and
a doubling of the unit cell.
While this transition was originally thought to be spin--Peierls,
recent studies have found evidence for charge
order.\cite{ohama,isobe,neutron}
It has been proposed that this material is well--described as a quarter-filled
ladder.\cite{Smol98,Seo98}
Two--leg ladder models are also thought to be
relevant to a number of other materials containing ladder--like structures,
such as the the Vanadates  MgV$_2$O$_5$ and CaV$_2$O$_5$, and the
cuprates SrCu$_2$O$_3$ and Sr$_{14}$Cu$_{24}$O$_{41}$.
For a more detailed description of ladder materials and models,
we direct the reader to a recent review
\cite{dagottoreview} and the references contained therein.

One of the simplest models of interacting electrons that allows for
charge ordering is the Hubbard model extended with
an additional nearest--neighbor (NN) Coulomb repulsion, $V$.
The charge order transition in this model has been studied in the
one-dimensional (1D) model in the strong--coupling
limit,\cite{hubbard78} at quarter filling
\cite{Penc94} and
at half filling,\cite{Cannon91,Zhang97} in the 2D system at half filling,
\cite{Zhang89,Chatto97} and within the Dynamical Mean Field Theory (the
limit of infinite dimensions) at quarter \cite{Pietig99}
and half filling. \cite{Dongen94}
A variety of techniques, such as mean-field approximations,
perturbation theory, and numerical methods as Quantum Monte Carlo and
the Density Matrix Renormalization Group (DMRG) have been employed in
these studies.
Their results can be summarized as follows:
At the mean-field level, the transition between a homogeneous
state and a charge-density wave (CDW) state at half filling in a
hypercubic lattice occurs at $V_c = U/z_0$ where
$z_0$ denotes the number of nearest neighbors ($z_0 = 2d$).
Numerical studies \cite{Zhang97} indicate a slightly higher value of $V_c$,
at least in 1D.
Interestingly, in 1D at half filling
the transition has been found to be second order
at small $U/t$ and first order at large $U/t$ with the tricritical
point located at $U_c/t \sim 4-6$. \cite{Zhang97}
Here we use the term ``first order'' to denote discontinuous behavior
of the charge order parameter as a function of microscopic
parameters such as $V$ or band filling, and ``second order'' to denote
continuous behavior.
At quarter filling in 1D, the situation is more complicated since
a number of phases compete for large $V$. \cite{Penc94}
In higher dimensions, the transition seems to always be first order.
\cite{Zhang89,Chatto97,Pietig99,Dongen94}
However, conclusive studies that can
reliably distinguish between first-- and second--order transitions are
lacking.
Between quarter and half filling, the existence and nature of a
transition is unclear in general.
In 1D, the extended Hubbard model undergoes phase separation rather
than a transition to a CDW state for $|U|/t < 4$ and large $V$. \cite{Clay99}
For $U/t > 4$, indications are that the 1D system undergoes a
transition to a $q=\pi$ CDW state for sufficiently large
$V$ at all fillings.\cite{Lin95}

In this paper, we examine the charge-order transition in the
extended Hubbard model on the two--leg ladder,
considering all values of band filling between quarter and half
filling.
We shall investigate the nature and location of the charge-order
transition in the ladder and compare with the 1D as well as higher
dimensional models.



The single--band extended Hubbard model has the Hamiltonian
\begin{eqnarray}
H&=&-t\sum_{\langle ij\rangle\sigma}(c^\dagger_{i\sigma} c_{j\sigma}^{} + h.c.)
\nonumber\\
 && + U\sum_{i} n_{i\uparrow} n_{i\downarrow}
    + V \sum_{\langle ij\rangle} n_i n_j \; .
\label{hamiltonian}
\end{eqnarray}
Here we consider a lattice consisting of two chains of length $L$,
i.e., a ladder, and discuss band fillings $\langle n \rangle = N/(2L)$,
with $N$ the number of electrons.
The summation $\langle ij\rangle$ then runs over all pairs of nearest
neighbor sites in the ladder.
In this work, we take both the hopping and the nearest-neighbor
Coulomb repulsion to be isotropic; in general, one could consider
the anisotropic case, parameterized by  $t_\parallel$, $t_\perp$,
$V_\parallel$ and $V_\perp$.

The numerical results shown in this paper have been calculated with
the DMRG technique \cite{DMRG} on lattices of up to $2\times 64$
sites with open boundary conditions both between the two chains and at
the ends of the chains.
We have kept up to 600 states per block, resulting in the discarded weight
of the density matrix eigenvalues being typically $10^{-8}$ or less.
The errors in the energies and correlation functions arising from the
truncation of the density matrix are always less than a few percent.



\begin{figure}
\epsfxsize=7.5 truecm
\centerline{\epsffile{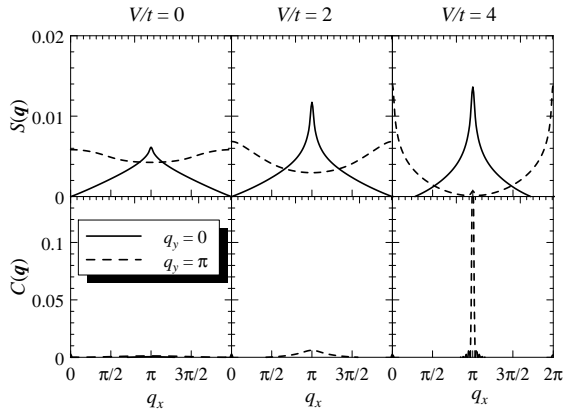}}
\caption{
  Charge- and spin correlation functions for a $64 \times 2$ ladder, $U=8 t$,
  quarter filling $\langle n\rangle=0.5$ (64 electrons),
  and different values of the nearest-neighbor repulsion $V$.
}
\label{FIG_CORR1}
\end{figure}

We now turn to the discussion of the results.
In order to investigate the charge ordering, we consider the
static charge structure factor
\begin{equation}
C({\bf q}) = {1 \over 2 L} \sum_{i}
  e^{i {\bf q} \cdot {\bf R}_i} \bar{C}({\bf R}_i)
\end{equation}
where
\begin{equation}
\bar{C}({\bf R}_i) = \frac{1}{N_{\rm av}}\sum_{\{j \}}
\langle \delta n_{j+i} \delta n_j\rangle  \; ,
\end{equation}
$\langle ... \rangle$ denotes the ground-state expectation value,
$\delta n_j = n_j - \langle n_j \rangle $, and we average over
typically $N_{\rm av} = 6$ sites to remove oscillations due to the open
boundaries.
For checkerboard charge order, which is expected to be the ground state
for large values of $V$, $C({\bf q})$ should show a pronounced
peak at ${\bf Q} = (\pi,\pi)$.
The spin order can be studied by looking at the spin structure factor,
S({\bf q}), which is defined similarly in terms of
the spin--spin correlation function $\langle S_{j+i}^z S_j^z \rangle$.

In Fig.\ \ref{FIG_CORR1}, we show the results for $C({\bf q})$
and $S({\bf q})$ for
a $2\times 64$ system at $U = 8 t$ and quarter filling for different values
of $V$.
The transition between a homogeneous state at small $V$ and
an ordered CDW state at large $V$ can clearly be seen.
The DMRG ground state at $V>V_c$ is fully gapped and has broken translation
invariance, i.e., for large $V$ every second site is empty.
[This gapped charge--ordered state with $(\pi,\pi)$ broken symmetry
is present at all fillings from quarter filling to half filling.]
In the spin channel, peaks at $(0,\pi)$ and $(\pi,0)$
develop with increasing $V$ showing an alternating spin pattern
on the {\it occupied} sites in the CDW state.
This spin order arises from virtual hopping processes (fourth order in
$t$) which lead to antiferromagnetic couplings between the
occupied sites.
In the limit of large $V$ the system can be
mapped to a $J_1-J_2$ spin-$1\over 2$ chain where $J_1 \gg J_2$ because of
the larger number of exchange processes in diagonal (1,1) direction.

\begin{figure}
\epsfxsize=7 truecm
\centerline{\epsffile{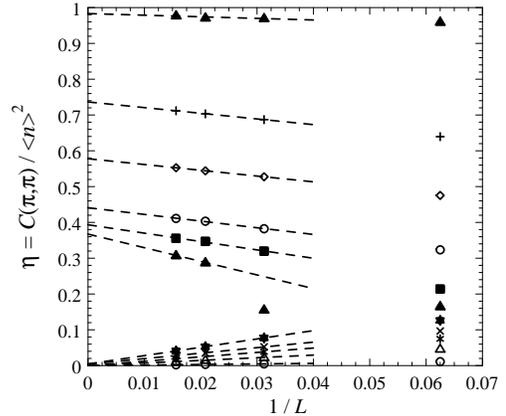}}
\caption{
  Finite-size scaling for the staggered charge
  correlation function at $U=8 t$ and $\langle n\rangle=0.75$.
  The curves correspond to $V/t=2.5,2.8,2.9,2.95,3.0,3.05,3.1,3.2,3.5,4.0,$
  and 8.0 from bottom to top.
  The dashed lines are linear fits through the data points for $L=32,48,64$.
  The irregular behavior at $V/t=3.05$ may indicate a first
  order transition (here the $L=32$ point was excluded from the fit).
}
\label{FIG_SCALE1}
\end{figure}

At half filling, the spin correlations correspond to short-range
$(\pi,\pi)$ order at small $V/U$.
In the limit of large $V \gg (t,U)$ the spin correlations vanish.
This behavior is due to the increase in the number of double occupancies
in the charge-ordered state at half filling, which leads to a
suppression of local magnetic moments.
Between quarter and half filling, $S({\bf q})$ shows peaks at
incommensurate wavevectors that correspond to combinations of bonding
and antibonding Fermi vectors \cite{noacktwochains}.
Near quarter filling, these peaks tend to be strengthened but also
displaced as $V$ is increased, whereas closer to half filling the
peaks tend to be suppressed in the CDW phase.
We further note that between quarter and half filling the residual magnetic
moments tend to order ferromagnetically in the charge-ordered phase,
i.e., the ground-state spin is non-zero. \cite{LadderPD}
The existence of gapless modes in the homogeneous phase is less
clear \cite{noacktwochains};
at least at half filling both spin and charge gaps are non-zero for
any parameters with $U>0$.

In order to examine the extent of the charge ordering and the nature
of the transition, we calculate the order parameter for a
${\bf Q} = (\pi,\pi)$ charge--ordered state,
\begin{equation}
 \eta = \lim_{L\rightarrow\infty} {C({\bf Q}) \over \langle n\rangle^2 } \; .
\end{equation}
We carry out the $L\rightarrow\infty$ extrapolation by calculating
$\eta(L)$ for $L=16,32,48,64$ and performing a linear fit in $1/L$
through the three largest system sizes, see Fig. \ref{FIG_SCALE1}.
(The $L=16$ point is consistent with the linear extrapolation except in
the region of the transition.)
This extrapolation is the major source of uncertainty in our results;
the error here is typically $\Delta\eta = 0.01$.

\begin{figure}
\epsfxsize=6 truecm
\centerline{\epsffile{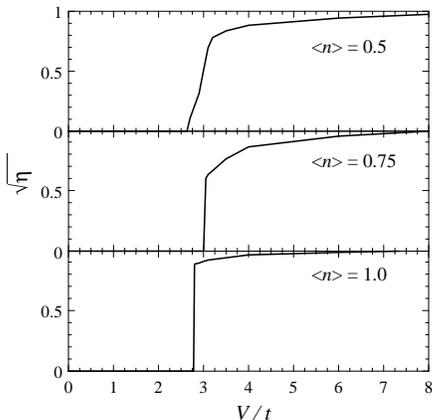}}
\caption{
  Order parameter $\sqrt{\eta}$ versus $V/t$ for $U=8 t$
  and $\langle n\rangle=0.5$, 0.75 and 1.
  The data indicate a second-order quantum phase transition at
  $V_c/t= 2.65 \pm 0.1$
  between a homogeneous and a charge-ordered state for $\langle n\rangle=0.5$.
  For $\langle n\rangle=0.75$ the transition at $V_c/t = 3.05 \pm 0.1$
  is weakly first order
  whereas the transition at half filling at $V_c/t = 2.8 \pm 0.1$ is
  clearly first order.
}
\label{FIG_OP1}
\end{figure}

In Fig. \ref{FIG_OP1}, we plot $\sqrt{\eta}$ (which
corresponds to the relative difference of the sublattice occupancies
in the broken-symmetry charge-ordered state) for $U=8 t$ and
different band fillings.
For $U=4t$ as well as $U=8t$, we find a transition
from a homogeneous state to a charge-ordered state with
increasing $V$ for all values of $\langle n \rangle$.
Note that the extrapolation with system size is crucial in order to
extract the $V_c$ from our data.
Examination of $\eta(V;L)$ for {\it fixed} $L$, as
used in Ref. \onlinecite{Zhang97} to determine $V_c$ from
DMRG results for the 1D system, is not adequate for the ladder system.

An additional, related method to determine $V_c$ is to examine the
scaling of $L \cdot C({\bf Q};L)$ with $L$.
This quantity should diverge with $L$ in the
charge-ordered state, but
should decrease with $L$ in the homogeneous state to a value determined by the
density correlation length of the system. \cite{Cannon91}
This method yields the same values of $V_c$ as that obtained from the
extrapolation of $\sqrt{\eta}$ to within $\Delta V_c \approx 0.1 t$.
Furthermore, we have checked that the compressibility of the system is
positive, i.e., there is no tendency to phase separation in the
parameter region studied here ($U/t \ge 4$).
Note, however, that phase separation occurs for smaller values of $U/t$, e.g.,
for $U/t = 1$ and $V \gg t$.
A complete investigation of the phase diagram will be published elsewhere.
\cite{LadderPD}

\begin{figure}
\epsfxsize=6.5 truecm
\centerline{\epsffile{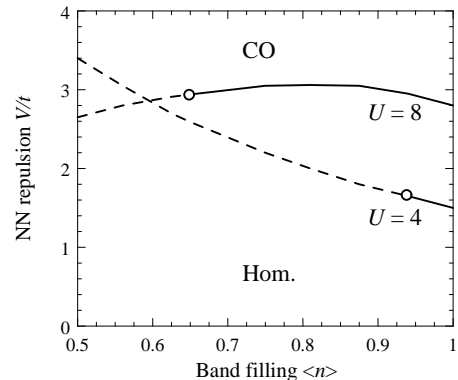}}
\caption{
  Phase diagram in the $V$--$\langle n\rangle$ plane
  showing first order (solid lines) and second order (dashed lines)
  transitions between
  the homogeneous and the charge-ordered state for $U=4 t$ and $U=8 t$.
  The boundaries have error $\Delta V_{c} \approx 0.1 t$ which
  arises mainly from
  the $L\rightarrow\infty$ extrapolation.
  The circles indicate the estimated positions of the tricritical
  point with uncertainty $\Delta \langle n_{tc}\rangle \approx 0.1$.
}
\label{FIG_PD1}
\end{figure}

At $U = 8t$ (Fig. \ref{FIG_OP1}) the
transition to the charge-ordered
state is continuous for small filling (e.g. quarter filling, $\langle
n \rangle=0.5$) whereas for large filling
the order parameter clearly shows a jump at a critical value $V_c$.
This leads to a tricritical point in the $V$--$\langle n \rangle$
plane, as can be seen in the phase diagram of Fig. \ref{FIG_PD1}.
This tricritical behavior seems to be similar to the one observed in
the 1D system at half filling. \cite{Cannon91,Zhang97}
The location of the tricritical point depends on the strength of
$U$, as shown in Fig. \ref{FIG_PD1}.
For $U=4 t$, its position is $\langle n \rangle_{tc} = 0.95 \pm 0.1$,
and at $U=8 t$ we find $\langle n \rangle_{tc} = 0.65 \pm 0.1$.
The position of the tricritical point (for fixed $U$) has been estimated from
the $\langle n \rangle$-dependence of the magnitude of the jump
in $\eta(V)$.

\begin{figure}
\epsfxsize=6.5 truecm
\centerline{\epsffile{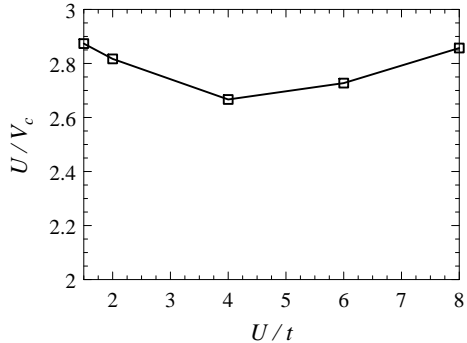}}
\caption{
  The ratio $U/V_c$ at half filling for different values of $U$.
  The data indicate that $U/V_c$ approaches the value $z_0 = 3$ for both
  weak and strong coupling which is consistent with mean-field
  arguments.
}
\label{FIG_VCU1}
\end{figure}

Finally, we address the dependence of $V_c$ on $U$ and the band
filling.
At half filling, both
weak- and strong-coupling approximations \cite{Dongen94,Hartree} for
the transition between spin-density wave (SDW) and CDW yield a critical value of
$V_c = U/z_0$ for a system on a hypercubic lattice.
However, these
Hartree-type approximations assume a SDW phase with true long-range
order which is absent for 1D chains as well as for the ladder
system considered here.
Nevertheless, the mean-field $V_c$
is in reasonable agreement with the value of slightly more than
$U/2$ found numerically for a 1D chain.\cite{Zhang97}
For the ladder systems studied in the present work, the DMRG results
for half filling also agree quite well with the predictions of the
Hartree approximation.
As can be seen in Fig. \ref{FIG_VCU1}, $U/V_c$ is consistent
with $V_c = U/3$ in both the weak- and strong-coupling
limits.

Below half filling, the behavior is more complicated because
it is no longer dominated only by the interplay of $U$ and $V$.
Since the average number of double occupancies decreases with
decreasing electron density, the kinetic energy becomes important
for larger $U$.
In the region between quarter filling and half filling, we
are faced with the interplay of all three energy scales $t$, $U$, and
$V$, and as can be seen in Fig.\ \ref{FIG_PD1}, $V_c$ varies considerably
with $\langle n \rangle$ and $U$.

In summary, our scenario for the charge-order transition in the
Hubbard ladder with $U \ge 4 t$ proceeds as follows:
At all band fillings, an increasing nearest-neighbor repulsion $V$
drives a transition from a homogeneous state to a gapped CDW state with
${\bf Q} = (\pi,\pi)$.
At quarter filling, each occupied site in the CDW state carries spin
$1 \over 2$, and the residual kinetic energy orders these spins
``antiferromagnetically''  at ordering vectors $(\pi,0)$ and $(0,\pi)$.
A similar charge ordering leading to ``zigzag'' antiferromagnetic ordering has
been found at quarter filling in coupled Hubbard ladders within
mean--field theory \cite{Seo98} and in
the $t$--$J$--$V$ ladder with adiabatic Holstein
phonons. \cite{RieraPoilblanc}
At half filling, the CDW state found at large $V$ consists of
doubly occupied sites without local moments, so that the spin order is
strongly suppressed.
The critical $V_c$ goes asymptotically to the
value $V_c = U/3$ in both the weak- and strong-coupling limits, as
predicted by mean--field theory.
Between quarter and half filling, there is a mixture of singly and
doubly occupied sites in the CDW phase,
and the spin order is, in general, incommensurate.
The total spin in the CDW phase is non-zero, i.e., charge order
co-exists with ferromagnetism.
As $V$ is increased and the system
goes into the CDW phase, the incommensurate spin correlations tend to be
enhanced closer to quarter filling and suppressed closer to half
filling.
Note that this suppression of the spin moments within the CDW phase as
the filling is increased is in contrast to
the behavior in the homogeneous phase at $V=0$ and $U \gg t$, for which the
spin moment {\it increases} with increasing $\langle n\rangle$.

An interesting subject for future study would be a more detailed comparison
of the phase diagram of the ladder system with that of
the 1D chain. \cite{Penc94,Clay99}
For instance, the phase separation region in the ladder for large $V/t$ is much
smaller \cite{LadderPD} than the $|U|/t < 4$ region found in 1D.
The origin of these differences in the charge-ordering process and the role of the
non-zero ground-state spin remain to be clarified.

The authors thank D.\ Baeriswyl, S.\ Blawid, R.\ Bulla and R.\ Pietig
for useful conversations.
M.V. acknowledges support by the DFG (VO 794/1-1) and R.M.N. support from
the Swiss National Foundation under Grant No 20--53800.98.
The calculations were performed on the Origin 2000 at the Technical University
Dresden.



\end{multicols}

\end{document}